# Two-dimensional charge density wave TaX$_2$ (X=S, Se, Te) from first principles

Tao Jiang[1], Tao Hu[1,2], Guodong Zhao[1], Yongchang Li[1], Shaowen Xu[1], Chao Liu[1], Yaning Cui[1] and Wei Ren[1,2,*]

[1] *Physics Department, International Center for Quantum and Molecular Structures, Shanghai Key Laboratory of High Temperature Superconductors, Shanghai University, Shanghai 200444, China*

[2] *State Key Laboratory of Advanced Special Steel, Materials Genome Institute, Shanghai University, Shanghai 200444, China*

* Email: renwei@shu.edu.cn

**ABSTRCT**

Transition metal dichalcogenides are rich in their structural phases, e.g. 1*T*-TaS$_2$ and 1*T*-TaSe$_2$ form charge density wave (CDW) under low temperature with interesting and exotic properties. Here, we present a systematic study of different structures in two-dimensional TaX$_2$ (X=S, Se, Te) using density functional theory calculations with consideration of van der Waals interaction. All the normal phases present metal characteristics with various ground state and magnetic properties. The lattice reconstruction of CDW drastically affects the electronic and structural characteristics of 1*T*-TaS$_2$ and 1*T*-TaSe$_2$, leading to a transition from metal to insulator and an emergence of magnetic moment within periodic atomic clusters called the Star of David. The evaluated Heisenberg couplings indicate the weak ferromagnetic coupling between the clusters in monolayer. Furthermore, in bilayer commensurate CDW cases, we find intriguing phenomenon of the varying magnetic properties with different stacking orders. The magnetic moment in each layer disappears when two layers are coupled, but may sustain in certain stackings of interlayer antiferromagnetic configurations.

## INTRODUCTION

Transition metal dichalcogenides (TMDs) have recently aroused much interest in their intriguing physical properties, ranging from magnetism, ferroelectricity to superconductivity [1-4]. Besides, TMDs are rich in presenting various structural phases such as 1*T*, 2*H*, 3*R* and *T'*, exhibiting different resulting phenomena. As a typical polymorphic TMD, TaS$_2$ has been studied extensively for its charge density wave (CDW) and superconductivity, as well as the competition between them [5-7]. Bulk 1*T*-TaS$_2$ undergoes a series of transitions upon decreasing the temperature. Below 550K, it transforms from a normal 1*T* phase to an incommensurate CDW phase; then enters a nearly commensurate CDW phase at 350K; and finally becomes a commensurate CDW(CCDW) phase at 180K, with 13 Ta atoms forming a "Star of David" cluster. Due to the dramatic increase of resistivity in CCDW compared with normal phase, the former is often classified as a Mott insulator because of the mysterious origin of the gap opening [8-9]. Through applying external modulation towards the transition between different CDW orders, including in-plane electric field, thickness engineering, gate voltage control, the CDW of 1*T*-TaS$_2$ shows prompting application in functional devices [10-15].

In two-dimensional (2D) case of TMDs, the properties of CDW show unique patterns different from their bulk cases. Recently, 2D 1$T$ and 2$H$ phases of TaS$_2$ and TaSe$_2$ have been successfully synthesized, from mechanical exfoliation to chemical vapor deposition and acid-assisted exfoliation [6-7, 13, 16-25]. 2D 1$T$-TaS$_2$ was found to develop a series of metastable states, which results from the slowing kinetics of phase transition when the thickness is reduced down to nanometer scale [26]. In addition, restacked 2$H$-TaS$_2$ nanosheets exhibit increased superconducting $T_c$ from 0.8K to 3K [18]. And the defects controlled by tuning pore density during chemical exfoliation of 2$H$-TaS$_2$ monolayer seems to enhance its superconductivity [6], contrary to the known fact that disorder hinders the superconductivity in 2D materials. Lian *et al.* [27] reported the coexistence of superconductivity and charge density wave in 2D 2H-TaSe$_2$ with its 3 × 3 CDW being the ground state. The CCDW of 2D 2$H$-TaS$_2$ at room temperature from heteroepitaxial growth on h-BN was discovered by using Raman spectra and scanning transmission electron microscopy [23]. However, in monolayer 1$T$-phase, the CCDW often emerges at low temperature, as described before. For the origin of the insulating ground state of 1$T$-TaS$_2$, there are two main points of view: one is Mott insulator [8] and the other is quantum spin liquid [28]. Recently, Lee *et al.* [29] proved that different stacking types of CDWs could also have an impact on the metal-insulator transition and insulating phase, in accompany with an interlayer Peierls dimerization. Then based on the observation of angle resolved photoemission spectroscopy (ARPES) and X-ray diffraction (XRD), the ground state of 1$T$-TaS$_2$ turned out to be a band insulator with interlayer dimerization, whereas a Mott insulator emerged between an extremely narrow temperature range [30]. What is more, a theoretical framework, based on the experimental research of 2D CDWs in 2$H$-MX$_2$ (X = S, Se and M = Nb, Ta), was proposed that ionic charge transfer, electron-phonon coupling and spatial extension of the electronic wave functions might have a comprehensive effect on the CDW formation in 2D materials [22]. However, little attention has been paid to the nature of CDW itself in the 2D limit, e.g. the appearance and absence of magnetism.

In this letter, we report systematic first-principles study of 2D TaX$_2$ (X=S, Se, Te) from their normal phases, including $T$, $H$ and $T'$, to the CCDW phase. For normal phases, the results show different ground state and magnetic properties of TaX$_2$. Then in $\sqrt{13} \times \sqrt{13}$ CCDW structure, we demonstrate its stability and a transition from metallic behavior to magnetic semiconductor in TaS$_2$ and TaSe$_2$. By constructing magnetic configurations, we obtain the Heisenberg exchange couplings which suggest a ferromagnetic ground state of CCDW phase of TaS$_2$ and TaSe$_2$. We further analyze the magnetic properties of bilayer CCDW phase of TaS$_2$ and TaSe$_2$ under different stacking orders. The magnetic properties of bilayer CCDW vary critically with the influence of staking patterns due to interlayer charger transfer. Our present work provides comprehensive investigation of TaX$_2$ that will shed light on the understanding of 2D CDWs.

## COMPUTATIONAL METHODS

The density functional theory (DFT) calculations were performed with the projector augment wave (PAW) [31] method implemented in the Vienna *ab initial* simulation package (VASP) [32-33]. Within the generalized gradient approximation (GGA) [34], we employed Perdew-Burke-Ernzerhof (PBE) exchange and correlation functional. The energy cutoff of 520 eV was chosen throughout all calculations. For the undistorted structure and $\sqrt{13} \times \sqrt{13}$ CCDW phase, $15 \times 15 \times 1$ and $6 \times 6 \times 1$ Γ-centered Monkhorst-Pack k meshes were used in geometry optimization and self-consistent calculations. The geometry optimization of the structures was achieved by converging all the forces and energies within 0.01 eV/Å and $10^{-6}$ eV, respectively. The van der Waals (vdW) correction [35] was adopted to reproduce the experimental lattice constants. To correct the band gap underestimation of the PBE functional, we further calculate the band structures of CCDW phase of 1$T$-TaX$_2$ (X=S, Se) using hybrid screened functional HSE06. In the calculation of magnetic properties of CCDW, an extra on-site Coulomb interaction was considered with effective $U = 2.70$ eV and 2.00 eV for TaS$_2$ [36-37] and TaSe$_2$ [38]. To prevent unphysical interaction between the periodic images, we employed a vacuum layer of at least 18 Å along c axis.

## RESULTS AND DISCUSSION
### I. UNDISTORTED PHASES

We present in Fig. 1 the monolayers of 2$H$ and 1$T$ TaX$_2$ (X=S, Se, Te), together with $T'$ phase, which are commonly seen in the family of TMDs. The $H$-TaX$_2$ has $D_{3h}$ point group and an X-Ta-X sandwich structure with Ta atoms arranged in trigonal prismatic coordination, while the $T$-phase belongs to $D_{3d}$ point group with octahedral coordination of Ta atoms. Both the $H$ and $T$ phases include one Ta and two X atoms in a unit cell, however, the $T'$-phase is distorted octahedral TMD having two Ta atoms in a unit cell, as shown in Fig. 1(c).

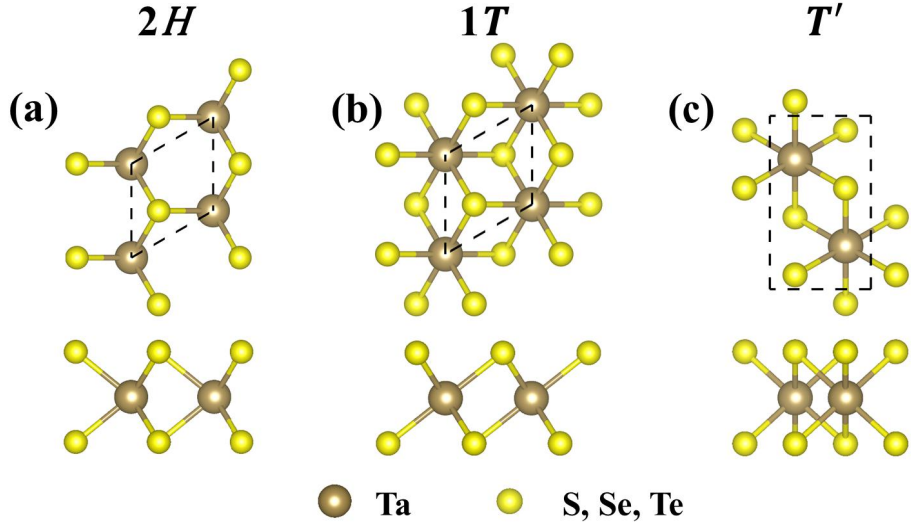

FIG. 1. The top and side views for (a) $H$-phase, (b) $T$-phase and (c) $T'$ phase of monolayer TaX$_2$ (X=S, Se, Te). The bigger brown spheres and smaller yellow spheres represent Ta and X atoms respectively. The black dashed lines indicate the unit cells.

To understand different phases of TaX$_2$, we first compare the ground state energy of undistorted phases based on fully optimized structures. From Table S1 (see Supporting Information), we can see that the most stable phase is $H$-phase for monolayer TaS$_2$ and TaSe$_2$ calculated by PBE+optB86-vdW, consistent with previous study [39-40]. Notably, the lattice parameters in our calculation are consistent with the experimental data [20, 38], thanks to the consideration of van der Waals interaction [39-40]. For the TaTe$_2$, we find the $T'$ phase is more stable and about 25 meV per formula unit (f.u.) lower in energy than the H-phase [40]. Moreover, all these three phases of TaX$_2$ monolayers show metallic behavior. But in terms of magnetic properties, only H-TaTe$_2$ shows a magnetic moment when PBE+optB86b is employed. We also double check the data using SCAN+rVV10 functional, which results non-zero moment for most of the H-TaX$_2$ (X=S, Se) as shown in Supporting Information Table S1. We choose PBE+optB86-vdW functional in the following calculations [41].

The electronic band structures (Supporting Information Fig. S1 ~ Fig. S3) show the absence of spin splitting in TaS$_2$ and TaSe$_2$ as other typical 2D-TMDs like MoS$_2$. However, the band splitting near Fermi level of H-TaTe$_2$ is distinctive, resulting a 0.14 $\mu_B$ magnetic moment per f.u., slightly smaller than previous calculation [40]. From the density of states (Supporting Information Fig. S1 ~ Fig. S3), Ta $d$ orbitals and some X $p$ orbitals dominate the states near the Fermi level. The magnetic moment of H-TaTe$_2$ is primarily contributed by Ta $d$ orbitals, about 0.128 $\mu_B$/Ta atom.

## II. CHARGE DENSITY WAVE 1T-PHASES
### A. MONOLAYER CCDW

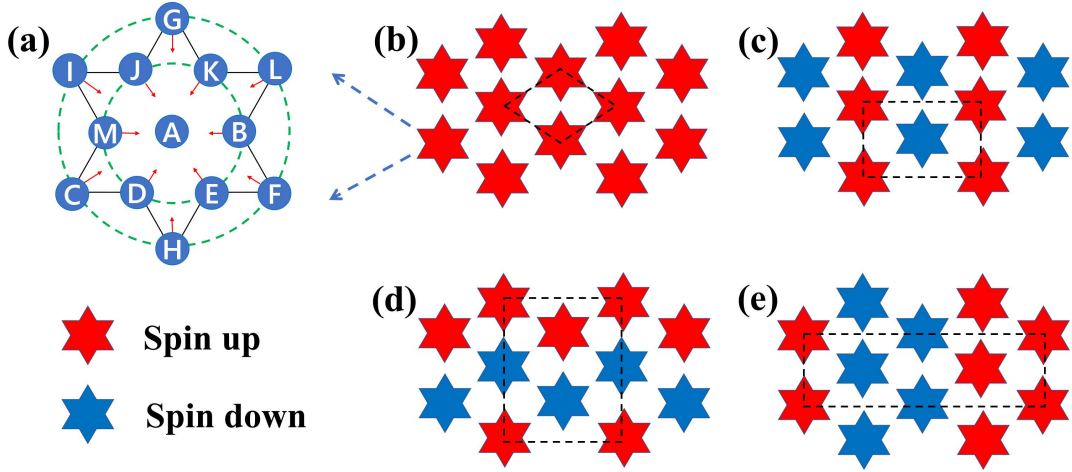

FIG. 2. Structure and magnetic configurations of the "Star of David" clusters. (a) The 13 sites of Ta atoms are labeled in one Star of David. (b) FM, (c) to (e) three AFM states. The red and blue stars correspond to the spin up and spin down states respectively. The black dashed lines represent the supercells except for (b) for a primitive cell.

By introducing initial displacements manually and relaxing atomic positions of the $\sqrt{13} \times \sqrt{13}$ supercell of 1T-TaX$_2$ (X=S, Se, Te), we obtained the CCDW phase as shown in Fig. 2, with one Ta atom in the center of two rings of six Ta atoms, namely the Star of David. Comparing the stability between CCDW and undistorted 1T-phase of monolayer TaS$_2$, we find that the lattice reconstruction can lower the energy by about 31 meV/f.u. (74 meV/f.u. for 1T-TaSe$_2$ and 106 meV/f.u. for 1T-TaTe$_2$) against undistorted phase, indicating that the CCDW phase is more stable than the normal 1$T$-phase.

Turning to the electronic structure of 1$T$-CCDW, we find that band structures of both TaS$_2$ and TaSe$_2$ present a magnetic semiconductor behavior, with band gaps about 0.088 eV and 0.051 eV respectively (the HSE gap are 0.609 eV and 0.473 eV) between two flat bands. To estimate experimental energy gap, we perform an extra GGA+$U$ band structure calculation to consider the electron correlation effect (Supporting Information Fig. S4). The energy band gaps for TaS$_2$ and TaSe$_2$ are found to be 0.213 eV and 0.153 eV respectively, consistent with the STM measurements [38]. Comparing the band structures of undistorted 1$T$ phases (Supporting Information Fig. S2) and CCDW phases, we confirm that the formation of CDW evidently leads to a gap opening, transforming the 1$T$-TaS$_2$ and 1$T$-TaSe$_2$ from a nonmagnetic metal to a magnetic semiconductor. Nevertheless, in the case of CCDW phase of 1$T$-TaTe$_2$, it remains a metal without any gap opening or band splitting (Fig. 3c).

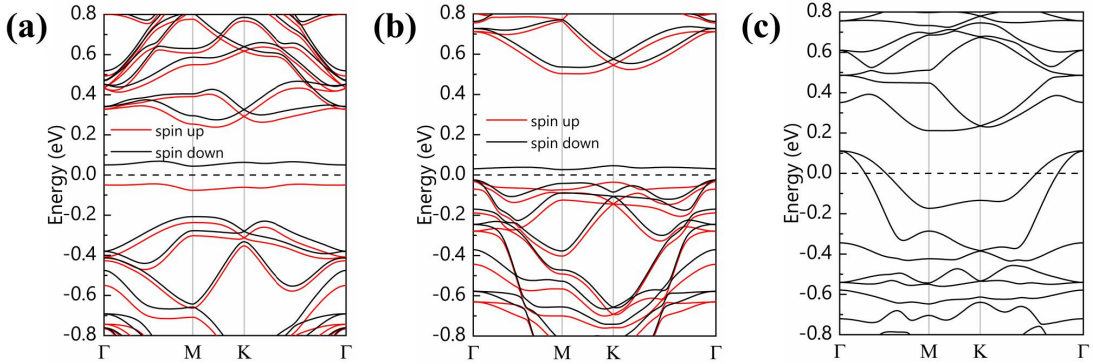

FIG. 3. The PBE band structures of CCDW phase of 1$T$-TaX$_2$. (a) TaS$_2$, (b) TaSe$_2$, (c) TaTe$_2$. The red and black solid lines indicate the spin up and down states respectively. The Fermi level is set to 0 eV by the black dashed line.

Supporting Information Fig. S5 shows the phonon dispersion of undistorted phase and CCDW phase of monolayer 1$T$-TaX$_2$ calculated by density perturbation theory method [42]. Experimental efforts have been made to report the various phases from 1$T$-TaS$_2$, and at low temperature (below 180 K) bulk 1$T$-TaS$_2$ should form a CCDW phase, while the ideal 1$T$-TaS$_2$ bulk shows unstable phonon dispersions [43]. This effect is also seen in our negative frequency phonon bands in the Brillouin zone for undistorted monolayers. Recently, 2D TaS$_2$ and TaSe$_2$ have been exfoliated experimentally to show phenomenon of CDW at low temperature [20, 38]. As expected, our phonon dispersion results of the CCDW phases do not present any imaginary frequency, in agreement with the experimental findings.

Next, we concentrate on the magnetic properties of CCDW phase of 1$T$-TaS$_2$ and 1$T$-TaSe$_2$. Both TaS$_2$ and TaSe$_2$ are found to have 1 $\mu_B$ moment per Star of David at GGA+$U$ level. Quantitatively the Ta $d$ orbital contributes majority of the total magnetic moment, and the central Ta atom approximately provides two fifth of total moment. Charge transfer from outer Ta atoms to the central one with the periodic lattice distortion, can be seen in the spin density in Fig. 4. To explore how neighboring Stars of David couple each other, here, we consider a ferromagnetic (FM) and three types of antiferromagnetic (AFM) configurations [44], as shown in Fig. 2. We give the relative energy of different magnetic configurations in Supporting Information Table S2, in which both TaS$_2$ and TaSe$_2$ have FM ground state.

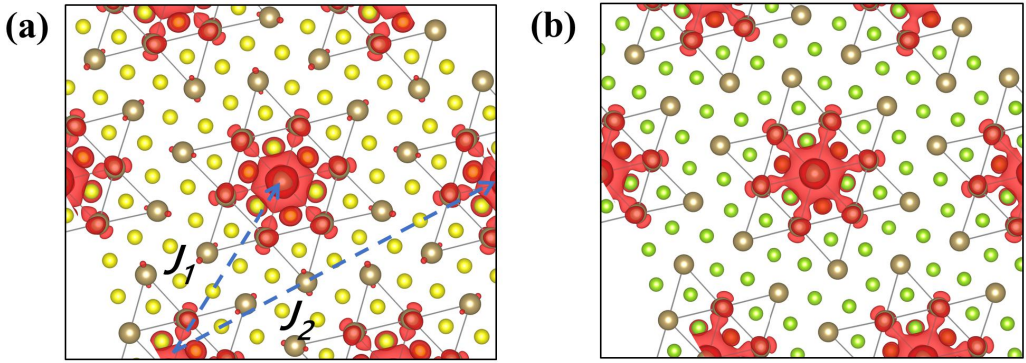

FIG. 4. Spin-polarization density of monolayer CCDW structure obtained at GGA+$U$ level of (a) 1$T$-TaS$_2$ and (b) 1$T$-TaSe$_2$. The isosurface value was set to 0.0015 $e/Bohr^3$. The exchange coupling parameters $J_1$ and $J_2$ are indicated by the blue dashed arrows.

According to the energy of different magnetic states, we can also deduce the Heisenberg exchange couplings (Table S2), given by $H = -\sum_{ij}J_1 S_i S_j - \sum_{ik}J_2 S_i S_j - A S_i^Z S_i^Z$, where $J_1$ and $J_2$ represent the nearest and next-nearest exchange couplings, $S_i$ is the spin vector of each Star of David clusters and $A$ is the anisotropy parameter. The $J_1$ values for TaS$_2$ and TaSe$_2$ are 0.341meV and 0.128 meV respectively, suggesting a rather weak nearest-neighbor ferromagnetic coupling, and the $J_2$ parameters have quite small negative values of -0.035 meV and -0.011 meV for TaS$_2$ and TaSe$_2$ respectively. By employing Monte Carlo simulations, we obtain the $T_c$ of FM to paramagnetic transition for TaS$_2$ and TaSe$_2$ are 6 K and 2 K, respectively.

### B. BILAYER CCDW AND STACKING ORDERS

In this section, we consider the bilayer CCDW of 1$T$-TaS$_2$ and 1$T$-TaSe$_2$ with different stacking patterns reported by Lee *et al* [29]. Taken the central Ta atom in bottom layer as reference A, the center of "Star of David" cluster in top layer can be arranged in 13 different ways, on the top of A, B, C, …, M, as shown in Fig. 2(a). However, there are only five distinct sites due to the threefold rotational symmetry, namely A, B, C, L and M. These interlayer stacking orders are labeled as AA, AB, AC, AL and AM respectively. In our calculations, we assign an initial magnetic moment for each Ta atom of the bilayer CCDW to generate its interlayer FM or AFM coupling. Based on the fully relaxed structures, we found that the stacking orders affect the magnetic properties of bilayer CCDW, like CrI$_3$ bilayers [45]. As shown in Supporting Information Table S3, the relative energy values of TaS$_2$ and TaSe$_2$ bilayers vary from 0 to ~60 meV/Star for different stackings. It is apparent that the AA stacking is the most stable configuration for both TaS$_2$ and TaSe$_2$, and the equilibrium interlayer spacing of AA is the largest among all stackings. As mentioned, monolayer CCDW phases of 1$T$-TaS$_2$ and 1$T$-TaSe$_2$ both have 1 $\mu_B$ moment per Star. However, in TaS$_2$ bilayer, the AC and AL have an interlayer AFM ground state, while the other three stacking orders all become nonmagnetic. In other words, each layer has the local magnetic moment slightly less than 1 $\mu_B$ per Star with opposite signs in the AC and AL stackings. The magnetic moments of all Ta atoms in both layers disappear in the other three stackings. As for TaSe$_2$, five stackings are all nonmagnetic, so that when two TaSe$_2$ monolayers bind together, the magnetic moment of each Ta atom becomes zero.

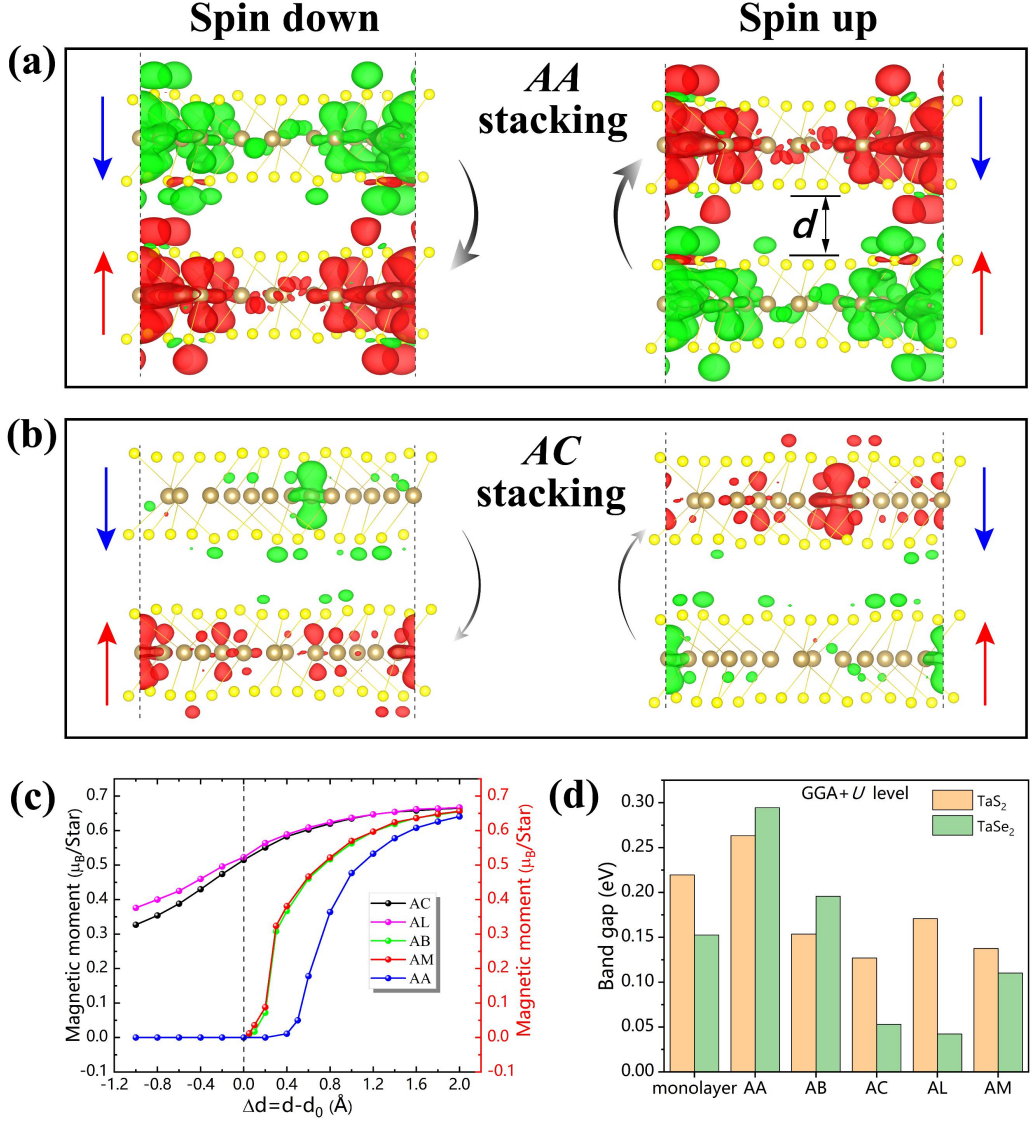

FIG. 5. Spin dependent interlayer differential charge density (SDCD) of TaS$_2$ bilayer. (a) Spin up and spin down SDCD of AA stacking. (b) Spin up and spin down SDCD of AC stacking. The red and green colors indicate the spin charge accumulation and depletion respectively. (c) The local magnetic moments as function of the change of interlayer distance where $d_0$ represents the equilibrium interlayer distance of the five stackings. (d) The band gaps for different stackings of TaS$_2$ and TaSe$_2$ CCDW.

To interpret this interesting phenomenon in bilayer CCDWs, we plot the spin dependent interlayer differential charge density (SDCD) [46] of bilayer TaS$_2$ in different stacking orders with an isosurface value of 0.0005 $e/Bohr^3$ in Fig. 5. With the top layer assigned with spin down and the bottom layer with spin up configurations, the SDCD of AA stacking shows significantly spin down charge transfer from the top layer to the bottom layer [left panel of Fig. 5(a)]. While the right panel of Fig. 5(a) shows the opposite case that the spin up charge transfer from bottom layer to the top layer. Since the spin charge transfer in Fig. 5(a) is equal to each other in magnitude in AA stacking (the same for AB and AM), the magnetic moments are cancelled out. The SDCD of interlayer AFM AC stacking [Fig. 5(b)] shows little spin charge transfer compared with AA stacking, which explains very slight decrease of magnetic moment in each layer from monolayer CCDW. The equilibrium interlayer distance of bilayer CCDW (about 2.73Å for AA) is significantly small compared with CrI$_3$ [45] and MoS$_2$ [47] bilayers. We thus expect that the interlayer distance affects the interlayer spin charge transfer and further modifies the magnetic properties. To confirm this, we change the interlayer distance of nonmagnetic stackings (AA, AB and AM) and AFM AC or AL stackings to examine the consequent magnetic configurations. As expected, when the interlayer distance is increased by 0.4 Å in AA stacking (0.1 Å for AB and AM stackings), the local magnetic moment emerges and approaches to the monolayer limit, as shown in Fig. 5(c), suggesting the elimination of spin charge transfer. While in AC and AL cases, the local magnetic moment of each Star decreases with the reduction of interlayer distance, which indicates the increase of interlayer charge transfer. The intralayer magnetic moment remains even if the layer spacing is reduced by 1 Å, highlighting the dependence of magnetic configurations on the stacking patterns.

To further confirm the impact of interlayer spin charge transfer on the magnetic configurations, we performed relative sliding of top layer in AA-TaS$_2$ along the diagonal direction as shown in Supporting Information Fig. S6(a). As discussed in section IIA, the magnetic moment in monolayer CCDW is mostly contributed by the central Ta atom. The distance between central Ta atoms from two layers (Supporting Information Table S3) shows that $d_{Ta-Ta}$ values of AFM stackings are larger than those of non-magnetic ones, thus hindering the extent of spin charge transfer. It should be noted that the $d_{Ta-Ta}$ values of AB and AM are slightly smaller than AA stacking, so that in Fig. 5 the critical $\Delta d$ for emerging magnetic moment of AB and AM is shorter than the AA. When the top layer of AA stacking is displaced along the diagonal to change the $d_{Ta-Ta}$, the magnetic moment emerges abruptly and reaches the peak value at one third of the diagonal (about 1 $\mu_B/Star$), and then diminishes with the reduction of $d_{Ta-Ta}$, as shown in Supporting Information Fig. S6. This result reconfirms the spin charge transfer in bilayer CCDW and explains the fascinating magnetic properties in different stacking orders.

Supporting Information Fig. S7 presents the electronic band structures of bilayer TaS$_2$ CCDWs at GGA+$U$ level. The most distinctive feature of the band structure with respect to monolayer is the spin degeneracy caused by interlayer interaction. Generally, the band gap reduces with the addition of layers due to the quantum confinement effect, e.g. MoS$_2$ [47] and black phosphorus few layers [48]. The variation of band gap shows fluctuation for different stacking configurations in TaS$_2$ and TaSe$_2$, as depicted in Fig. 5(d). The band gap of monolayer TaS$_2$ CCDW is smaller than the AA bilayer but larger than the other four stackings, while that for TaSe$_2$ the monolayer is smaller than AA and AB but larger than the other three stackings. This is consistent with the few-layer CCDW TaSe$_2$ reports of STM experiment which presented the energy band gap narrowing from single-layer to trilayer of AC stacking [38]. Our GGA+*U* simulations provide an overview of the influence of interlayer interaction on the energy gap in different stacking orders. Thus, we propose that the evolution of band gaps of CCDW TaS$_2$ and TaSe$_2$ few layers might be tailored by the stacking patterns.

## CONCLUSIONS

In summary, we investigated the structural, electronic and magnetic properties of normal phase TaX$_2$ (X=S, Se, Te) and *T* phase CCDW in two dimensions. The normal phases of TaS$_2$ and TaSe$_2$ are both non-magnetic metals with *H*-phase as the ground state, while TaTe$_2$ *H*-phase being a magnetic metal and *T″* phase a non-magnetic metal as the ground state. Then we focus on the monolayer $\sqrt{13} \times \sqrt{13}$ CCDW structure. Our calculations show that the lattice reconstruction turns the *T*-TaS$_2$ and *T*-TaSe$_2$ to magnetic semiconductor and become dynamically stable in contrast to the normal *T*-phase. Through three AFM and one FM magnetic configurations, our obtained Heisenberg exchange constants suggest the weak nearest-neighbor FM coupling between the Stars of David in TaS$_2$ and TaSe$_2$ monolayers. In the end, we established bilayer CCDWs with five different stacking orders which interestingly demonstrate their impact on the interlayer charge transfer and magnetic properties. We hope such systematic exploration of different phases, especially for CCDW structures, may help better understand the tantalum metal dichalcogenides and other TMDs. These findings in few layer CCDWs will enrich the physics of 2D-CDWs and stimulate more experimental and theoretical works of *T*-TaS$_2$ and *T*-TaSe$_2$ for their potential applications in spintronics.

## ACKNOWLEDGEMENTS

This work was supported by the National Natural Science Foundation of China (51861145315, 11929401, 12074241, 11804216), the Independent Research and Development Project of State Key Laboratory of Advanced Special Steel, Shanghai Key Laboratory of Advanced Ferrometallurgy, Shanghai University (SKLASS 2020-Z07), the Science and Technology Commission of Shanghai Municipality (19DZ2270200, 19010500500, 20501130600), and High Performance Computing Center, Shanghai University.

(1) Liu, X.; Pyatakov, A. P.; Ren, W. Magnetoelectric Coupling in Multiferroic Bilayer VS$_2$. *Physical Review Letters* **2020**, *125* (24), 247601,

(2) Liu, X.; Yang, Y.; Hu, T.; Zhao, G.; Chen, C.; Ren, W. Vertical ferroelectric switching by in-plane sliding of two-dimensional bilayer WTe$_2$. *Nanoscale* **2019,** *11* (40), 18575-18581,

(3) Hu, T.; Jia, F.; Zhao, G.; Wu, J.; Stroppa, A.; Ren, W. Intrinsic and anisotropic Rashba spin splitting in Janus transition-metal dichalcogenide monolayers. *Physical Review B* **2018,** *97* (23), 235404,

(4) Wang, L.; Shih, E. M.; Ghiotto, A.; Xian, L.; Rhodes, D. A.; Tan, C.; Claassen, M.; Kennes, D. M.; Bai, Y. S.; Kim, B.; Watanabe, K.; Taniguchi, T.; Zhu, X. Y.; Hone, J.; Rubio, A.; Pasupathy, A. N.; Dean, C. R. Correlated electronic phases in twisted bilayer transition metal dichalcogenides. *Nature Materials* **2020,** *19* (8), 861-866,

(5) Sipos, B.; Kusmartseva, A. F.; Akrap, A.; Berger, H.; Forro, L.; Tutis, E. From Mott state to superconductivity in 1T-TaS2. *Nature Materials* **2008,** *7* (12), 960-965,

(6) Peng, J.; Yu, Z.; Wu, J. J.; Zhou, Y.; Guo, Y. Q.; Li, Z. J.; Zhao, J. Y.; Wu, C. Z.; Xie, Y. Disorder Enhanced Superconductivity toward TaS$_2$ Monolayer. *ACS Nano* **2018,** *12* (9), 9461-9466,

(7) Yang, Y. F.; Fang, S.; Fatemi, V.; Ruhman, J.; Navarro-Moratalla, E.; Watanabe, K.; Taniguchi, T.; Kaxiras, E.; Jarillo-Herrero, P. Enhanced superconductivity upon weakening of charge density wave transport in 2H-TaS$_2$ in the two-dimensional limit. *Physical Review B* **2018,** *98* (3), 035203,

(8) Fazekas, P.; Tosatti, E. Electrical, structural and magnetic properties of pure and doped 1T-TaS$_2$. *Philosophical Magazine B* **1979,** *39* (3), 229-244,

(9) Rossnagel, K.; Smith, N. V. Spin-orbit coupling in the band structure of reconstructed 1T−TaS$_2$. *Physical Review B* **2006,** *73* (7), 073106,

(10) Hollander, M. J.; Liu, Y.; Lu, W. J.; Li, L. J.; Sun, Y. P.; Robinson, J. A.; Datta, S. Electrically driven reversible insulator-metal phase transition in 1T-TaS$_2$. *Nano Letters* **2015,** *15* (3), 1861-6,

(11) Yu, Y. J.; Yang, F. Y.; Lu, X. F.; Yan, Y. J.; Cho, Y. H.; Ma, L. G.; Niu, X. H.; Kim, S.; Son, Y. W.; Feng, D. L.; Li, S. Y.; Cheong, S. W.; Chen, X. H.; Zhang, Y. B. Gate-tunable phase transitions in thin flakes of 1T-TaS$_2$. *Nature Nanotechnology* **2015,** *10* (3), 270-276,

(12) Li, L.; Zhang, J.; Myeong, G.; Shin, W.; Lim, H.; Kim, B.; Kim, S.; Jin, T.; Cavill, S.; Kim, B. S.; Kim, C.; Lischner, J.; Ferreira, A.; Cho, S. Gate-Tunable Reversible Rashba-Edelstein Effect in a Few-Layer Graphene/2H-TaS$_2$ Heterostructure at Room Temperature. *ACS Nano* **2020,** *14* (5), 5251-5259,

(13) Husain, S.; Chen, X.; Gupta, R.; Behera, N.; Kumar, P.; Edvinsson, T.; Garcia-Sanchez, F.; Brucas, R.; Chaudhary, S.; Sanyal, B.; Svedlindh, P.; Kumar, A. Large Damping-Like Spin-Orbit Torque in a 2D Conductive 1T-TaS$_2$ Monolayer. *Nano Letters* **2020,** *20* (9), 6372-6380,

(14) Liu, G.; Debnath, B.; Pope, T. R.; Salguero, T. T.; Lake, R. K.; Balandin, A. A. A charge-density-wave oscillator based on an integrated tantalum disulfide-boron nitride-graphene device operating at room temperature. *Nature Nanotechnology* **2016,** *11* (10), 845-850,

(15) Zhu, C.; Chen, Y.; Liu, F.; Zheng, S.; Li, X.; Chaturvedi, A.; Zhou, J.; Fu, Q.; He, Y.; Zeng, Q.; Fan, H. J.; Zhang, H.; Liu, W. J.; Yu, T.; Liu, Z. Light-Tunable 1T-TaS$_2$ Charge-Density-Wave Oscillators. *ACS Nano* **2018,** *12* (11), 11203-11210,

(16) Li, H.; Tan, Y.; Liu, P.; Guo, C.; Luo, M.; Han, J.; Lin, T.; Huang, F.; Chen, M. Atomic-Sized Pores Enhanced Electrocatalysis of TaS$_2$ Nanosheets for Hydrogen Evolution. *Advanced Materials* **2016,** *28* (40), 8945-8949,

(17) Albertini, O. R.; Zhao, R.; McCann, R. L.; Feng, S.; Terrones, M.; Freericks, J. K.; Robinson, J. A.; Liu, A. Y. Zone-center phonons of bulk, few-layer, and monolayer 1T-TaS$_2$: Detection of commensurate charge density wave phase through Raman scattering. *Physical Review B* **2016,** *93* (21), 214109,

(18) Pan, J.; Guo, C.; Song, C.; Lai, X.; Li, H.; Zhao, W.; Zhang, H.; Mu, G.; Bu, K.; Lin, T.; Xie, X.; Chen, M.; Huang, F. Enhanced Superconductivity in Restacked TaS$_2$ Nanosheets. *Journal of the American Chemical Society* **2017,** *139* (13), 4623-4626,

(19) Huan, Y.; Shi, J.; Zou, X.; Gong, Y.; Zhang, Z.; Li, M.; Zhao, L.; Xu, R.; Jiang, S.; Zhou, X.; Hong, M.; Xie, C.; Li, H.; Lang, X.; Zhang, Q.; Gu, L.; Yan, X.; Zhang, Y. Vertical 1T-TaS$_2$ Synthesis on Nanoporous Gold for High-Performance Electrocatalytic Applications. *Advanced Materials* **2018,** *30* (15), e1705916,

(20) Wang, X.; Liu, H.; Wu, J.; Lin, J.; He, W.; Wang, H.; Shi, X.; Suenaga, K.; Xie, L. Chemical Growth of 1T-TaS$_2$ Monolayer and Thin Films: Robust Charge Density Wave Transitions and High Bolometric Responsivity. *Advanced Materials* **2018,** *30* (38), 1800074,

(21) Wu, J.; Peng, J.; Yu, Z.; Zhou, Y.; Guo, Y.; Li, Z.; Lin, Y.; Ruan, K.; Wu, C.; Xie, Y. Acid-Assisted Exfoliation toward Metallic Sub-nanopore TaS$_2$ Monolayer with High Volumetric Capacitance. *Journal of the American Chemical Society* **2018,** *140* (1), 493-498,

(22) Lin, D.; Li, S.; Wen, J.; Berger, H.; Forro, L.; Zhou, H.; Jia, S.; Taniguchi, T.; Watanabe, K.; Xi, X.; Bahramy, M. S. Patterns and driving forces of dimensionality-dependent charge density waves in 2H-type transition metal dichalcogenides. *Nature Communications* **2020,** *11* (1), 2406,

(23) Fu, W.; Qiao, J.; Zhao, X.; Chen, Y.; Fu, D.; Yu, W.; Leng, K.; Song, P.; Chen, Z.; Yu, T.; Pennycook, S. J.; Quek, S. Y.; Loh, K. P. Room Temperature Commensurate Charge Density Wave on Epitaxially Grown Bilayer 2H-Tantalum Sulfide on Hexagonal Boron Nitride. *ACS Nano* **2020,** *14* (4), 3917-3926,

(24) Li, H.; Liu, P.; Liu, Q.; Luo, R.; Guo, C.; Wang, Z.; Guan, P.; Florencio Aleman, C.; Huang, F.; Chen, M. Twisted 1T TaS$_2$ bilayers by lithiation exfoliation. *Nanoscale* **2020,** *12* (35), 18031-18038,

(25) Tsen, A. W.; Hovden, R.; Wang, D.; Kim, Y. D.; Okamoto, J.; Spoth, K. A.; Liu, Y.; Lu, W.; Sun, Y.; Hone, J. C.; Kourkoutis, L. F.; Kim, P.; Pasupathy, A. N. Structure and control of charge density waves in two-dimensional 1T-TaS$_2$. *Proceedings of the National Academy of Sciences of the United States of America* **2015,** *112* (49), 15054-9,

(26) Masaro Yoshida, R. S. Memristive phase switching in two-dimensional 1T-TaS$_2$ crystals. *Science advances* **2015,** *1*, e1500606,

(27) Lian, C. S.; Heil, C.; Liu, X.; Si, C.; Giustino, F.; Duan, W. Coexistence of Superconductivity with Enhanced Charge Density Wave Order in the Two-Dimensional Limit of $TaSe_2$. *The Journal of Physical Chemistry Letters* **2019,** *10* (14), 4076-4081,

(28) Law, K. T.; Lee, P. A. 1T-TaS2 as a quantum spin liquid. *Proceedings of the National Academy of Sciences of the United States of America* **2017,** *114* (27), 6996-7000,

(29) Lee, S. H.; Goh, J. S.; Cho, D. Origin of the Insulating Phase and First-Order Metal-Insulator Transition in 1T-$TaS_2$. *Physical Review Letters* **2019,** *122* (10), 106404,

(30) Wang, Y. D.; Yao, W. L.; Xin, Z. M.; Han, T. T.; Wang, Z. G.; Chen, L.; Cai, C.; Li, Y.; Zhang, Y. Band insulator to Mott insulator transition in 1T-$TaS_2$. *Nature Communications* **2020,** *11* (1), 4215,

(31) Blochl, P. E. Projector augmented-wave method. *Physical Review B* **1994,** *50* (24), 17953-17979,

(32) Kresse, G. Efficient iterative schemes for ab initio total-energy calculations using a plane-wave basis set. *Physical Review B* **1996,** *54*, 169-186,

(33) Kresse, G.; Furthmuller, J. Efficiency of ab-initio total energy calculations for metals and semiconductors using a plane-wave basis set. *Computational Materials Science* **1996,** *6* (1), 15-50,

(34) Perdew, J. P.; Burke, K.; Ernzerhof, M. Generalized gradient approximation made simple. *Physical Review Letters* **1996,** *77* (18), 3865-3868,

(35) Klimes, J.; Bowler, D. R.; Michaelides, A. Van der Waals density functionals applied to solids. *Physical Review B* **2011,** *83* (19), 195131,

(36) Darancet, P.; Millis, A. J.; Marianetti, C. A. Three-dimensional metallic and two-dimensional insulating behavior in octahedral tantalum dichalcogenides. *Physical Review B* **2014,** *90* (4), 045134,

(37) Gan, L. Y.; Zhang, L. H.; Zhang, Q.; Guo, C. S.; Schwingenschlogl, U.; Zhao, Y. Strain tuning of the charge density wave in monolayer and bilayer 1T-$TaS_2$. *Physical Chemistry Chemical Physics* **2016,** *18* (4), 3080-5,

(38) Chen, Y.; Ruan, W.; Wu, M.; Tang, S.; Ryu, H.; Tsai, H.-Z.; Lee, R.; Kahn, S.; Liou, F.; Jia, C.; Albertini, O. R.; Xiong, H.; Jia, T.; Liu, Z.; Sobota, J. A.; Liu, A. Y.; Moore, J. E.; Shen, Z.-X.; Louie, S. G.; Mo, S.-K.; Crommie, M. F. Strong correlations and orbital texture in single-layer 1T-$TaSe_2$. *Nature Physics* **2020,** *16* (2), 218-224,

(39) Ding, Y.; Wang, Y.; Ni, J.; Shi, L.; Shi, S.; Tang, W. First principles study of structural, vibrational and electronic properties of graphene-like $MX_2$ (M=Mo, Nb, W, Ta; X=S, Se, Te) monolayers. *Physica B: Condensed Matter* **2011,** *406* (11), 2254-2260,

(40) Guo, H.; Lu, N.; Wang, L.; Wu, X.; Zeng, X. C. Tuning Electronic and Magnetic Properties of Early Transition-Metal Dichalcogenides via Tensile Strain. *The Journal of Physical Chemistry C* **2014,** *118* (13), 7242-7249,

(41) Fu, Y.; Singh, D. J. Applicability of the Strongly Constrained and Appropriately Normed Density Functional to Transition-Metal Magnetism. *Physical Review Letters* **2018,** *121* (20), 207201,

(42) Baroni, S.; de Gironcoli, S.; Dal Corso, A.; Giannozzi, P. Phonons and related crystal properties from density-functional perturbation theory. *Reviews of Modern Physics* **2001,** *73* (2), 515-562,

(43) Lazar, P.; Martincová, J.; Otyepka, M. Structure, dynamical stability, and electronic properties of phases in TaS2 from a high-level quantum mechanical calculation. *Physical Review B* **2015,** *92* (22), 224104,

(44) Wu, Q.; Zhang, Y.; Zhou, Q.; Wang, J.; Zeng, X. C. Transition-Metal Dihydride Monolayers: A New Family of Two-Dimensional Ferromagnetic Materials with Intrinsic Room-Temperature Half-Metallicity. *The Journal of Physical Chemistry Letters* **2018,** *9* (15), 4260-4266,

(45) Sivadas, N.; Okamoto, S.; Xu, X.; Fennie, C. J.; Xiao, D. Stacking-Dependent Magnetism in Bilayer CrI3. *Nano Letters* **2018,** *18* (12), 7658-7664,

(46) Wang, C.; Zhou, X.; Zhou, L.; Pan, Y.; Lu, Z.-Y.; Wan, X.; Wang, X.; Ji, W. Bethe-Slater-curve-like behavior and interlayer spin-exchange coupling mechanisms in two-dimensional magnetic bilayers. *Physical Review B* **2020,** *102* (2), 020402,

(47) Rudren Ganatra, Q. Z. Few-Layer $MoS_2$: A Promising Layered Semiconductor. *ACS Nano* **2014,** *8*, 4074–4099,

(48) Tran, V.; Soklaski, R.; Liang, Y. F.; Yang, L. Layer-controlled band gap and anisotropic excitons in few-layer black phosphorus. *Physical Review B* **2014,** *89* (23), 235319,

# Supporting Information for

# Two-dimensional charge density wave TaX$_2$ (X=S, Se, Te) from first principles

Tao Jiang[1], Tao Hu[1,2], Guodong Zhao[1], Yongchang Li[1], Shaowen Xu[1], Chao Liu[1], Yaning Cui[1] and Wei Ren[1*]

[1] *Physics Department, International Center for Quantum and Molecular Structures, Shanghai Key Laboratory of High Temperature Superconductors, Shanghai University, Shanghai 200444, China*

[2] *State Key Laboratory of Advanced Special Steel, Materials Genome Institute, Shanghai University, Shanghai 200444, China*

* Email: renwei@shu.edu.cn

TABLE S1. Lattice parameters of three phases of TaX$_2$, relative energy ΔE and magnetic moment per TaX$_2$ unit cell calculated by using PBE and SCAN functionals with van der Waals correction.

|  | TaS$_2$ | | | TaSe$_2$ | | | TaTe$_2$ | | |
| --- | --- | --- | --- | --- | --- | --- | --- | --- | --- |
|  | 1T | 2H | T' | 1T | 2H | T' | 1T | 2H | T' |
| PBE+vdW | | | | | | | | | |
| a(Å) | 3.351 | 3.318 | 5.805 | 3.471 | 3.448 | 6.013 | 3.646 | 3.673 | 6.567 |
| b(Å) | 3.351 | 3.318 | 3.350 | 3.471 | 3.448 | 3.473 | 3.646 | 3.673 | 3.380 |
| ΔE(meV/f.u.) | 78.307 | 0.000 | 76.930 | 81.436 | 0.000 | 77.860 | 6.689 | 0.000 | -25.010 |
| M ($\mu_B$/f.u.) | 0.000 | 0.000 | 0.000 | 0.000 | 0.000 | 0.000 | 0.000 | 0.140 | 0.000 |
| SCAN+vdW | | | | | | | | | |
| a(Å) | 3.354 | 3.319 | 5.813 | 3.474 | 3.448 | 6.014 | 3.620 | 3.660 | 6.532 |
| b(Å) | 3.354 | 3.319 | 3.354 | 3.474 | 3.448 | 3.476 | 3.620 | 3.660 | 3.342 |
| ΔE(meV/f.u.) | 103.119 | 0.000 | 101.280 | 139.112 | 0.000 | 135.872 | 70.668 | 0.000 | 2.102 |
| M ($\mu_B$/f.u.) | 0.000 | 0.345 | 0.000 | 0.297 | 0.452 | 0.475 | 0.084 | 0.519 | 0.000 |

TABLE S2. Relative energy (meV/Star) for different magnetic states of TaS$_2$ and TaSe$_2$ in CCDW phase.

| **Magnetic configurations** | **TaS$_2$ (meV/Star)** | **TaSe$_2$ (meV/Star)** |
| --- | --- | --- |
| FM | 0 | 0 |
| AFM1 | 1.255 | 0.469 |
| AFM2 | 1.134 | 0.492 |
| AFM3 | 0.341 | 0.128 |

TABLE S3. The relative energy $\Delta E$, equilibrium interlayer spacing $d$, distance between central Ta atoms in each layer $d_{Ta-Ta}$ and magnetic configurations of TaS$_2$ and TaSe$_2$ for various CDW stacking patterns.

| Stacking | $\Delta E$ (meV/Star) | $d$ (Å) | $d_{Ta-Ta}$ (Å) | Magnetic configuration |
|---|---|---|---|---|
| TaS$_2$ | | | | |
| AA | 0.000 | 2.725 | 6.037 | None |
| AB | 34.372 | 2.672 | 6.842 | None |
| AL | 34.929 | 2.684 | **8.338** | AFM |
| AM | 39.692 | 2.680 | 6.846 | None |
| AC | 58.989 | 2.688 | **8.320** | AFM |
| TaSe$_2$ | | | | |
| AA | 0.000 | 2.839 | 6.430 | None |
| AB | 25.589 | 2.731 | 7.239 | None |
| AM | 31.887 | 2.714 | 7.244 | None |
| AL | 51.225 | 2.709 | 8.717 | None |
| AC | 64.648 | 2.735 | 8.739 | None |

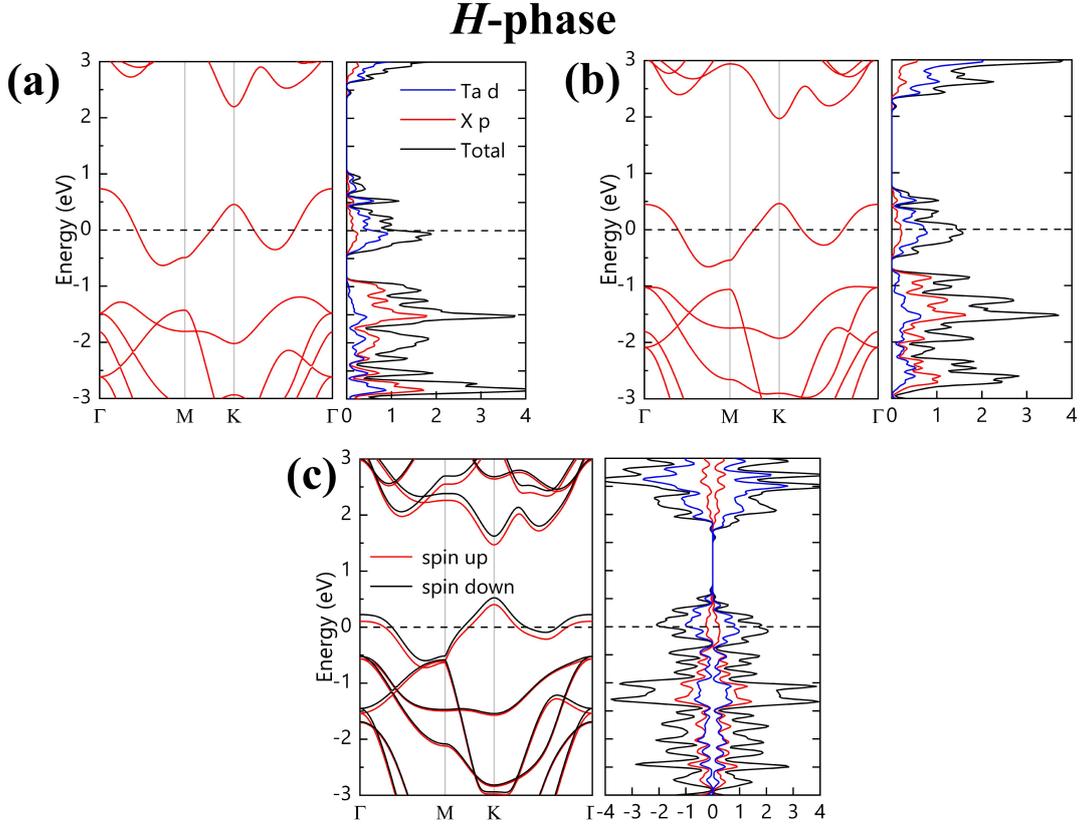

FIG. S1. Band structure and partial density of states of monolayer $H$-TaX$_2$ (X=S, Se, Te). (a) TaS$_2$, (b) TaSe$_2$ and (c) TaTe$_2$.

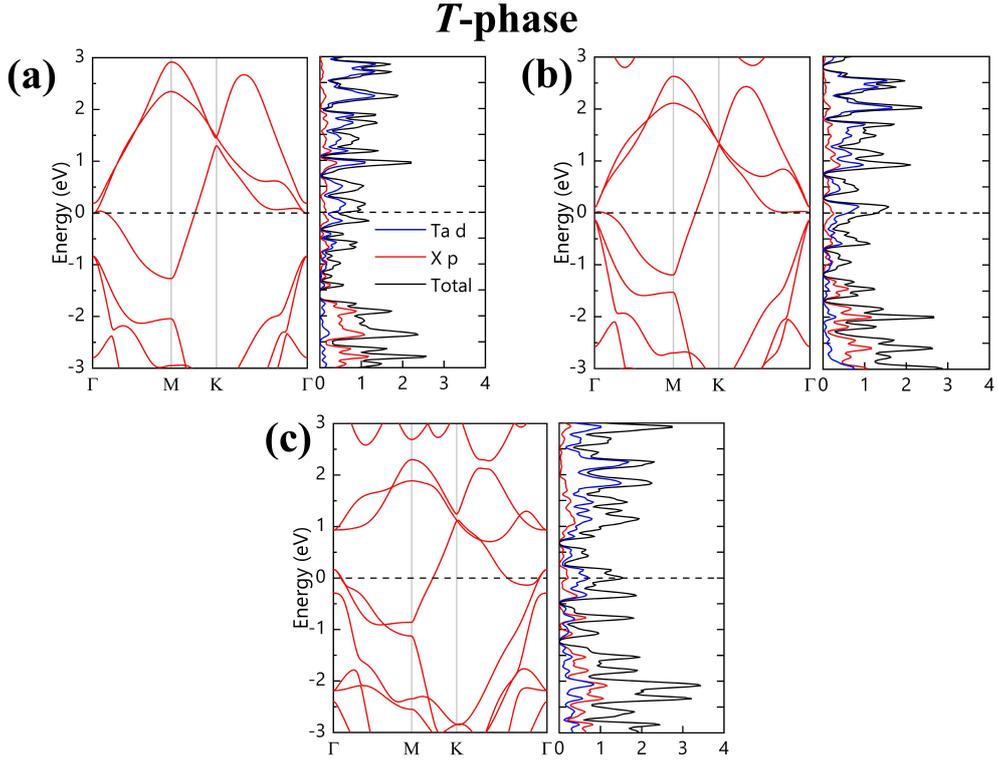

FIG. S2. Band structure and partial density of states of monolayer $T$-TaX$_2$ (X=S, Se, Te). (a) TaS$_2$, (b) TaSe$_2$ and (c) TaTe$_2$.

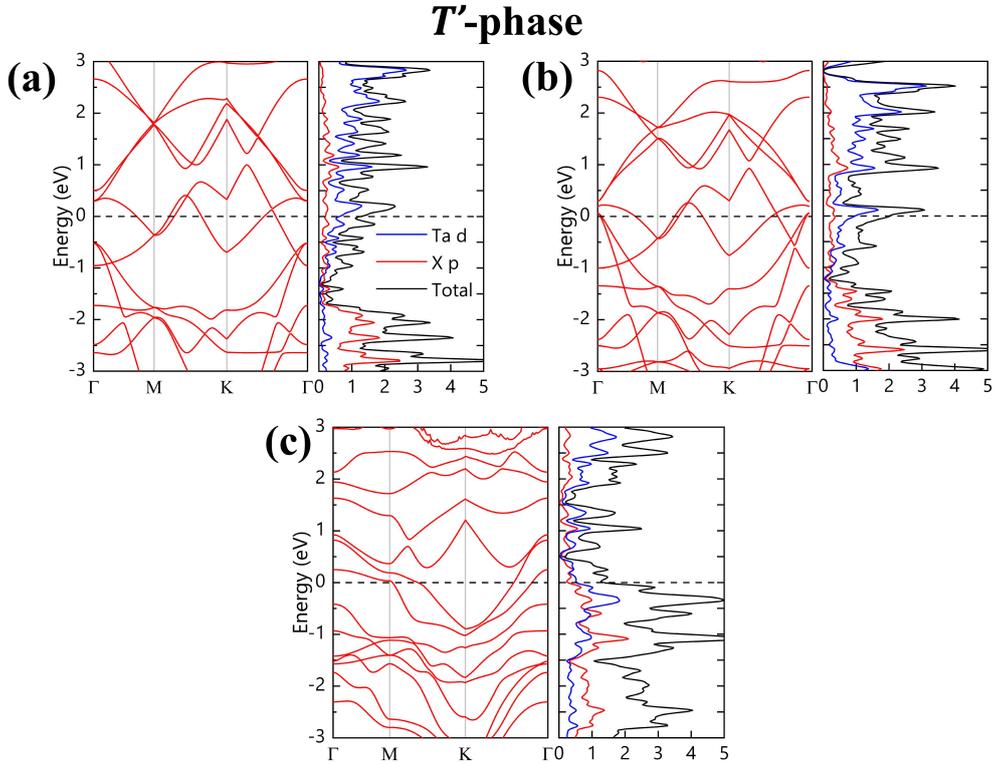

FIG. S3. Band structure and partial density of states of monolayer $T'$-TaX$_2$ (X=S, Se, Te). (a) TaS$_2$, (b) TaSe$_2$ and (c) TaTe$_2$.

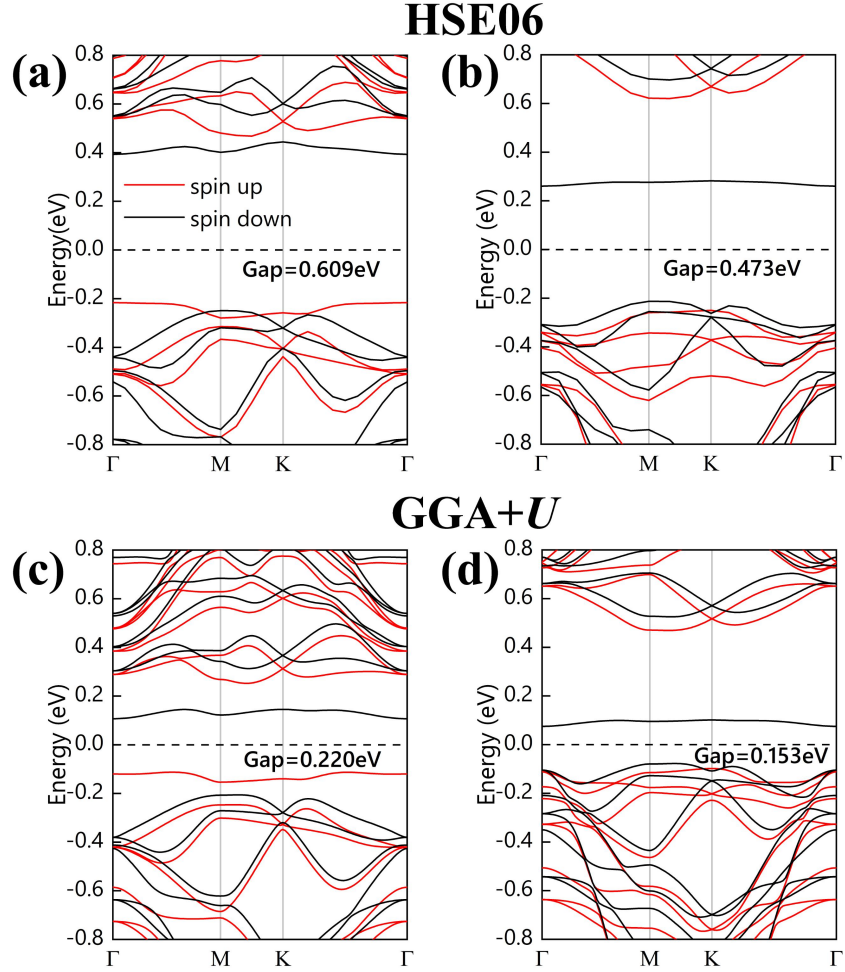

FIG. S4. Band structures of monolayer CCDW structures of $T$-TaS$_2$ (a) and (c); $T$-TaSe$_2$ (b) and (d) calculated by using HSE06 and GGA+$U$ functionals respectively.

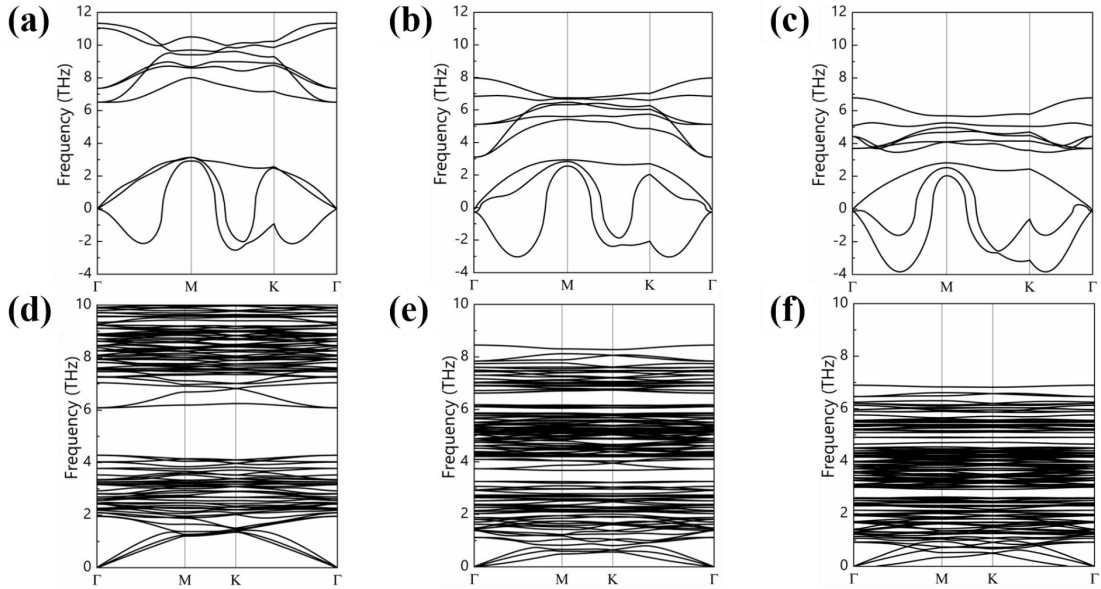

FIG. S5. Phonon dispersions of the undistorted (upper panels) and CCDW (lower) phases of monolayer 1$T$-TaS$_2$ (a) and (d); 1$T$-TaSe$_2$ (b) and (e); 1$T$-TaTe$_2$ (c) and (f).

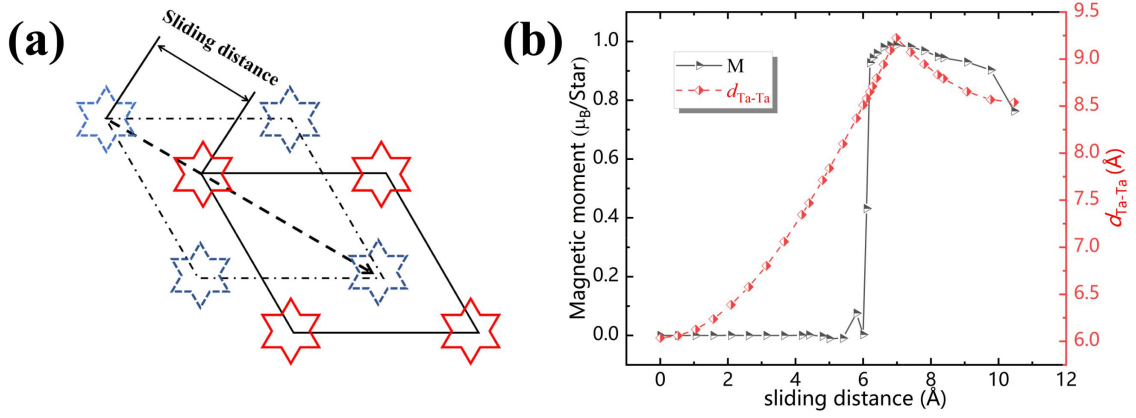

FIG. S6. (a) Illustration of the top layer sliding of bilayer CCDW TaS$_2$ from AA stacking to the one third of the diagonal. (b) The magnetic moment and $d_{Ta-Ta}$ as function of the sliding distance from AA to the half of the diagonal. The diagonal black dashed arrow in (a) indicates the sliding direction, the red solid and blue dashed stars represent the positions after and before sliding.

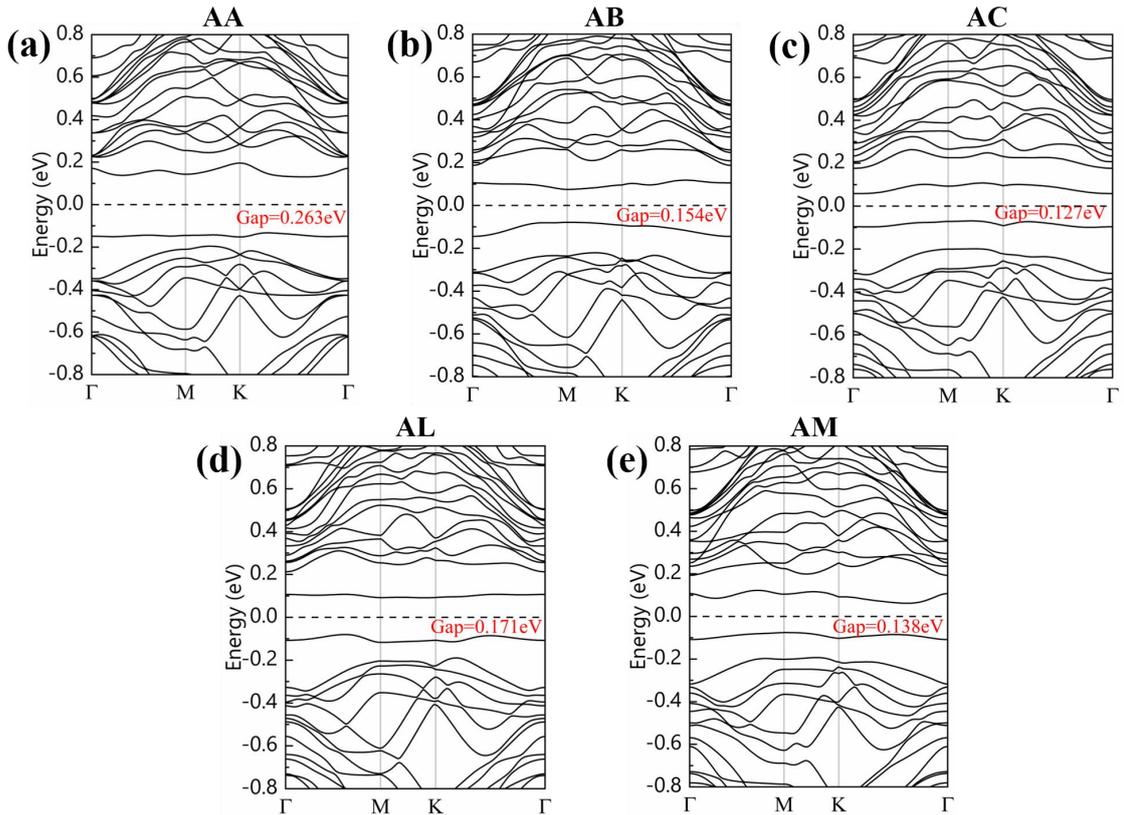

FIG. S7. GGA+$U$ ($U$ = 2.7eV) electronic band structures of five TaS$_2$ bilayers of different stacking patterns. The Fermi level is set to 0 eV by black dashed lines.